\documentclass
[showpacs,preprintnumbers,floats,twocolumn,10pt,prb,a4paper,floatfix]{revtex4}%
\usepackage{graphicx}
\usepackage{amsmath}
\usepackage{amsfonts}
\usepackage{amssymb}%
\begin{document}
\preprint{ }
\title{Escape rate of metastable states in a driven NbN superconducting microwave resonator}
\author{Baleegh Abdo}
\email{baleegh@tx.technion.ac.il}
\author{Eran Arbel-Segev}
\author{Oleg Shtempluck}
\author{Eyal Buks}
\affiliation{Department of Electrical Engineering, Technion, Haifa 32000, Israel}
\date{\today}

\begin{abstract}
We study thermal instability and formation of local hot spots in a driven
nonlinear NbN superconducting microwave resonator. White noise injected into
the resonator results in transitions between the metastable states via a
process consisting of two stages. In the first stage, the input noise entering
the system induces fluctuations in the resonator mode. While, in the second
one, these mode fluctuations result in phase transitions of the hot spot due
to induced temperature fluctuations. The associated noise-activated escape
rate is calculated theoretically, and also measured experimentally by means of
driving the system into stochastic resonance. A comparison between theory and
experiment yields a partial agreement.

\end{abstract}
\pacs{74.40.+k, 02.50.Ey, 85.25.-j }
\maketitle




\section{INTRODUCTION}

The simple model of noise activated escape of a Brownian particle over a
potential barrier successfully explains the basic behavior of a large number
of metastable systems in nature \cite{KramerRev}. Examples of such systems can
be found in almost all major fields of science: physics, chemistry, biology
and even engineering \cite{UsagesBook}. For instance, it explains biochemical
reactions in ac-driven protein \cite{protein}, the lifetime of zero-voltage
state in Josephson junctions \cite{Lifetime JJ,Thermal activation JJ}, the
magnetization reversal in nanomagnets
\cite{nanomagnetNeel,nanomagnetSubmicron,nanomagnetSpin}, noise-activated
switching in micro- \cite{micromechanical} and nano- \cite{Cleland,nano
Mohanty} mechanical oscillators, and photon-assisted tunneling in
semiconductor hetrostructures \cite{photon assisted}.

A well-known pioneering work on the subject is Krammer's in 1940. In his
seminal paper \cite{Kramer}, he derived relatively simple expressions for the
thermally induced escape rate in a one-dimensional asymmetric double-well
potential. In general, these escape rate expressions take the form of
$\Gamma=\Gamma_{0}\exp(-U_{b}/k_{B}T),$ where $U_{b}$ is the potential barrier
height, $k_{B}$ is Boltzmann's constant, $T$ is the temperature (where the
limit $k_{B}T\ll U_{b}$ is assumed), and $\Gamma_{0}$ is a rate prefactor.
Important extensions and refinements to this formula aimed either to include a
wider range of damping regimes \cite{overdamped,Exact solution,underdamped} or
accommodate the solutions to other cases such as nonequilibrium systems, have
been contributed by many authors over the years \cite{Graham,unequil
SIAM,unequil Ryv}. Examples of such nonequilibrium systems are metastable
potentials modulated by deterministic forces \cite{Path-integral} e. g. the
case of stochastic resonance \cite{high freq SR,SR review}, or metastable
systems subjected to nonwhite noise \cite{colored noise,Einchcomb}. Moreover,
efforts have been invested also in extending Krammer's rate theory to describe
metastable systems in the quantum limit \cite{KramerRev,Hanggi Q,tunneling
resonant structures}, where escape is dominated by tunneling.

In the present paper we study the escape rate of metastable states of
thermally instable superconducting stripline resonators both theoretically and
experimentally. In recent studies \cite{observation,dynamics} we have
experimentally demonstrated such instability in NbN superconducting
resonators. The measured response of the system to a monochromatic excitation
was successfully accounted for by a theoretical model, which attributed the
instability to a local hot spot in the resonator, switching between the
superconducting and the normal phases. Nonlinearity, according to this model,
results due to coupling between the equations of motion for both, the mode
amplitude in the resonator and the temperature of the hot spot. The coupling
mechanism is based on the dependence of both the resonance frequency and the
damping rate of the resonator on the stripline impedance, which in turn
depends on the temperature of the hot spot. Moreover, we have employed this
instability to demonstrate experimentally intermodulation gain \cite{IMgain},
stochastic resonance \cite{SRB}, self-sustained modulation of a monochromatic
drive \cite{Segev short,Segev long}, period doubling bifurcation, and noise
squeezing \cite{Segev extreme}.

In the case of thermally instable superconducting stripline resonators, the
escape mechanism governing the lifetime of the metastable states differs in
general from many of the examples mentioned above. In this case, the input
noise induces escape in a two-stage process. The direct coupling between the
input noise and the driven mode leads to fluctuations in the mode amplitude,
which in turn, induce fluctuations in the heating power applied to the hot
spot. Consequently, the fluctuating heating power, which is characterized by a
finite correlation time, leads to temperature fluctuations. Escape occurs when
the temperature approaches the critical value and a phase transition takes
place in the hot spot.

The remainder of this paper is organized as follows. In Sec. II the steady
state solutions of the equation of motion for the resonator-mode are derived
for the case of local heating instability. In Sec. III a perturbative approach
is applied in order to include the effect of thermal fluctuations. In Sec. IV
an escape rate expression characterizing the metastable states of the
resonator is obtained. In Sec. V a brief explanation regarding stochastic
resonance measurement is given. While in Secs. V. A and V. B stochastic
resonance measurement results are employed in order to extract some of the
transition rate parameters characterizing the system. Finally, a brief summary
concludes this paper in Sec VI.

\section{STEADY STATE SOLUTIONS}

Consider the case of a superconducting stripline microwave resonator weakly
coupled to a feedline. Driving the resonator by a coherent tone $a_{1}%
^{in}=b^{in}e^{-i\omega_{p}t}$ injected into the feedline, excites a mode in
the resonator with an amplitude $A=Be^{-i\omega_{p}t}$, where $\omega_{p}$ is
the drive angular frequency, $b^{in}$ is a constant complex amplitude
proportional to the drive strength, and $B\left(  t\right)  $ is a complex
mode amplitude which is assumed to vary slowly on the time scale of
$1/\omega_{p}$.

\subsection{Mode Amplitude}

In this approximation, the equation of motion for $B$ reads \cite{Yurke 05}%

\begin{equation}
\frac{\mathrm{d}B}{\mathrm{d}t}=\left[  i\left(  \omega_{p}-\omega_{0}\right)
-\gamma\right]  B-i\sqrt{2\gamma_{1}}b^{in}+c^{in}\;, \label{dB/dt}%
\end{equation}
where $\omega_{0}$ is the angular resonance frequency, $\gamma=\gamma
_{1}+\gamma_{2}$, where $\gamma_{1}$ is the coupling factor between the
resonator and the feedline, and $\gamma_{2}$ is the damping rate of the mode.
The term $c^{in}$ represents an input noise with a random phase%

\begin{equation}
\langle c^{in}\rangle=0\;, \label{<c>}%
\end{equation}
and autocorrelation functions given by%

\begin{equation}
\langle c^{in}(t)c^{in}(t^{\prime})\rangle=\langle c^{in\ast}(t)c^{in\ast
}(t^{\prime})\rangle=0\;, \label{<cc>=<c+c+>=0}%
\end{equation}

\begin{equation}
\langle c^{in}(t)c^{in\ast}(t^{\prime})\rangle=G\omega_{0}\delta\left(
t-t^{\prime}\right)  \;. \label{corr}%
\end{equation}

By further assuming a thermal equilibrium condition at temperature
$T_{\mathrm{eff}}$ and a relatively high temperature case $k_{B}%
T_{\mathrm{eff}}\gg\hbar\omega_{0}$, one has%

\begin{equation}
G=\frac{\gamma}{\omega_{0}}\frac{k_{B}T_{\mathrm{eff}}}{\hbar\omega_{0}}\;.
\label{G}%
\end{equation}

Rewriting Eq. (\ref{dB/dt}) in terms of the dimensionless time $\tau
=\omega_{0}t$ and using the steady state solution
\begin{equation}
B_{\infty}=\frac{i\sqrt{2\gamma_{1}}b^{in}}{i\left(  \omega_{p}-\omega
_{0}\right)  -\gamma}\;, \label{Binf}%
\end{equation}
yields the following compact form
\begin{equation}
\frac{db}{d\tau}+\lambda b=\frac{c^{in}}{\omega_{0}}\;, \label{db/d_tau}%
\end{equation}
where $b=B-B_{\infty}$ represents the difference between the mode amplitude
variable and the steady state solution, while $\lambda$ reads%

\begin{equation}
\lambda=\frac{\gamma-i\left(  \omega_{p}-\omega_{0}\right)  }{\omega_{0}}\;.
\end{equation}

By applying the methods of Gardiner and Collett introduced in Ref.
\cite{Gardiner} one can obtain the following input-output relation%

\begin{equation}
b^{out}=b^{in}-i\sqrt{2\gamma_{1}}B\;, \label{in out}%
\end{equation}
which relates the output signal $a_{1}^{out}=$ $b^{out}e^{i\omega_{p}t}$
reflected off the resonator to the input signal $a_{1}^{in}=b^{in}%
e^{-i\omega_{p}t}$ entering the system.

Thus, the reflection parameter $r$ in steady state is in general given by%

\begin{equation}
r=\frac{b^{out}}{b^{in}}=\frac{\gamma_{2}-\gamma_{1}-i\left(  \omega
_{p}-\omega_{0}\right)  }{\gamma_{2}+\gamma_{1}-i\left(  \omega_{p}-\omega
_{0}\right)  }\;, \label{r}%
\end{equation}
which is obtained by substituting $B_{\infty}$ of Eq. (\ref{Binf}) in the
input-output relation given by Eq. (\ref{in out}) and dividing by the input
drive amplitude $b^{in}$.

\subsection{Heat Balance of Local Heating}

The total power dissipated in the resonator $Q_{t}$ is given by
\begin{equation}
Q_{t}=\hslash\omega_{0}2\gamma_{2}\left\vert B\right\vert ^{2}\;.
\end{equation}
Furthermore, assuming that the resonator nonlinearity is dominated by a local
hot spot in the stripline resonator, and that the hot spot area is
sufficiently small in order to consider its temperature $T$ to be homogeneous,
the heat balance equation reads \cite{hot spots}%

\begin{equation}
C\frac{\mathrm{d}T}{\mathrm{d}t}=Q-W\;, \label{dT/dt}%
\end{equation}
where $C$ is the thermal heat capacity, $Q$ is the power heating up the hot
spot given by $Q=\alpha Q_{t},$ where $0\leqslant\alpha\leqslant1$, and
$W=H\left(  T-T_{0}\right)  $ is the power of the heat transfer to the
coolant, which is assumed to be at temperature $T_{0}$, where $H$ is the heat
transfer coefficient.

In terms of the dimensionless time $\tau$ and the dimensionless temperature
given by%

\begin{equation}
\Theta=\frac{T-T_{0}}{T_{c}-T_{0}}\;,
\end{equation}
Eq. (\ref{dT/dt}) reads%

\begin{equation}
\frac{\mathrm{d}\Theta}{\mathrm{d}\tau}+g\left(  \Theta-\Theta_{\infty
}\right)  =0\;, \label{dT/d_tau}%
\end{equation}
where the following quantities have been defined%

\begin{equation}
g=\frac{H}{C\omega_{0}}\;, \label{g}%
\end{equation}
and%

\begin{equation}
\Theta_{\infty}=\frac{2\hslash\alpha\gamma_{2}\left\vert B\right\vert ^{2}%
}{gC\left(  T_{c}-T_{0}\right)  }\;.
\end{equation}
Hence, the steady state solution of Eq. (\ref{dT/d_tau}), reads%

\begin{equation}
\Theta_{\infty0}=\frac{2\hslash\alpha\gamma_{2}\left\vert B_{\infty
}\right\vert ^{2}}{gC\left(  T_{c}-T_{0}\right)  }\;. \label{Theta_inf}%
\end{equation}

Moreover, if one further assumes that the fluctuation of $B$ around
$B_{\infty}$, is relatively small, one can rewrite Eq. (\ref{dT/d_tau}) in the
following form%

\begin{equation}
\frac{\mathrm{d}\theta}{\mathrm{d}\tau}+g\theta=f\;, \label{d theta /d tau}%
\end{equation}
where%

\begin{equation}
\theta=\Theta-\Theta_{\infty0}\;,
\end{equation}
and $f$ reads%

\begin{equation}
f=g\Theta_{\infty0}\left[  \frac{b}{B_{\infty}}+\left(  \frac{b}{B_{\infty}%
}\right)  ^{\ast}\right]  \;. \label{f}%
\end{equation}

In general, when a hot spot is generated or alternatively diminished in the
stripline, it affects the resonator parameters $\omega_{0}$, $\gamma_{1}$,
$\gamma_{2}$, $\alpha$ and may induce as a result jumps in the resonance
response curve. Moreover, as we have already shown in previous publications
\cite{dynamics,observation}, most of the nonlinear experimental results
exhibited by our superconducting NbN resonators can be modeled to a very good
extent by assuming a step function dependence of the resonator parameters
$\omega_{0}$, $\gamma_{1}$, $\gamma_{2}$, $\alpha$ on the hot spot temperature%

\begin{equation}
\omega_{0}=\left\{
\begin{array}
[c]{cc}%
\omega_{0s} & \Theta<1\\
\omega_{0n} & \Theta>1
\end{array}
\right.  \;,\;\gamma_{1}=\left\{
\begin{array}
[c]{cc}%
\gamma_{1s} & \Theta<1\\
\gamma_{1n} & \Theta>1
\end{array}
\right.  \;,
\end{equation}

\begin{equation}
\gamma_{2}=\left\{
\begin{array}
[c]{cc}%
\gamma_{2s} & \Theta<1\\
\gamma_{2n} & \Theta>1
\end{array}
\right.  \;,\;\alpha=\left\{
\begin{array}
[c]{cc}%
\alpha_{s} & \Theta<1\\
\alpha_{n} & \Theta>1
\end{array}
\right.  \;.
\end{equation}

In addition, we have shown that, in general, while disregarding noise, the
coupled equations (\ref{db/d_tau}) and (\ref{dT/d_tau}) may have up to two
different steady state solutions. A superconducting steady state (S) exists
when $\Theta_{\infty0}<1$, or when $E<E_{s}$, where $E_{s}=gC\left(
T_{c}-T_{0}\right)  /2\alpha_{s}\gamma_{2s}\hslash$. Similarly, a normal
steady state (N) exists when $\Theta_{\infty0}>1$, or when $E>E_{n}$, where
$E_{n}=gC\left(  T_{c}-T_{0}\right)  /2\alpha_{n}\gamma_{2n}\hslash$.

\section{FLUCTUATIONS}

In this section we assume a nonzero noise term $c^{in}\left(  t\right)  $
entering the resonator, thus giving rise to fluctuations around the steady
state solution.

\subsection{Mode Fluctuations}

In this case the solution of Eq. (\ref{db/d_tau}) reads%

\begin{equation}
b\left(  \tau\right)  =b\left(  0\right)  e^{-\lambda\tau}+\frac{1}{\omega
_{0}}\int_{0}^{\tau}c^{in}\left(  \tau^{\prime}\right)  e^{\lambda\left(
\tau^{\prime}-\tau\right)  }\mathrm{d}\tau^{\prime}\;.
\end{equation}

For relatively long times $\gamma\tau/\omega_{0}\gg1$ one gets by using Eq.
(\ref{<c>}) a zero mean value of the mode fluctuation $b$%

\begin{equation}
\left\langle b\left(  \tau\right)  \right\rangle =0\;,
\end{equation}
whereas by using Eqs. (\ref{<cc>=<c+c+>=0}), and (\ref{corr}) respectively,
one obtains the following autocorrelation functions%

\begin{equation}
\left\langle b\left(  \tau_{1}\right)  b\left(  \tau_{2}\right)  \right\rangle
=\left\langle b^{\ast}\left(  \tau_{1}\right)  b^{\ast}\left(  \tau
_{2}\right)  \right\rangle =0\;, \label{bb}%
\end{equation}
and%

\begin{align}
\left\langle b\left(  \tau_{1}\right)  b^{\ast}\left(  \tau_{2}\right)
\right\rangle  &  =\frac{G\omega_{0}}{2\gamma}e^{-\lambda^{\ast}\left\vert
\tau_{2}-\tau_{1}\right\vert }\;.\nonumber\\
&  \label{bbstar}%
\end{align}

\subsection{Local Heating Fluctuations}

Similarly, the solution of Eq. (\ref{d theta /d tau}) reads%

\begin{equation}
\theta\left(  \tau\right)  =\left\langle \theta\left(  \tau\right)
\right\rangle +\Delta_{\theta}\left(  \tau\right)  \;, \label{theta(tau)}%
\end{equation}
where%

\begin{equation}
\left\langle \theta\left(  \tau\right)  \right\rangle =\theta\left(  0\right)
e^{-g\tau}\;, \label{<theta>}%
\end{equation}
is the mean value of $\theta$ variable and%

\begin{equation}
\Delta_{\theta}\left(  \tau\right)  =\int_{0}^{\tau}f\left(  \tau^{\prime
}\right)  e^{g\left(  \tau^{\prime}-\tau\right)  }\mathrm{d}\tau^{\prime}\;,
\label{thetaDeviation}%
\end{equation}
is the deviation.

The variance of $\theta$, which is denoted as $\left\langle \Delta_{\theta
}^{2}\left(  \tau\right)  \right\rangle $, can be derived with the use of Eqs.
(\ref{thetaDeviation}), (\ref{f}), and (\ref{bbstar}). In the case of small
$\tau$, namely the case when $g\tau\ll1$ and $\left\vert \lambda\right\vert
\tau\ll1$, one has to lowest order in $\tau$%

\begin{equation}
\left\langle \Delta_{\theta}^{2}\left(  \tau\right)  \right\rangle
=\frac{g^{2}\Theta_{\infty0}^{2}}{\left\vert B_{\infty}\right\vert ^{2}}%
\frac{2G\omega_{0}\tau^{2}}{\gamma}\;. \label{Delta theta tau<<1}%
\end{equation}
On the other hand, for relatively long times $g\tau\gg1$ one finds%

\begin{align}
\left\langle \Delta_{\theta}^{2}\right\rangle  &  =\frac{G\Theta_{\infty0}%
^{2}}{\left\vert B_{\infty}\right\vert ^{2}}\frac{g\omega_{0}^{2}}{\gamma
}\frac{\gamma+g\omega_{0}}{\left(  \gamma+g\omega_{0}\right)  ^{2}+\left(
\omega_{p}-\omega_{0}\right)  ^{2}}\;.\nonumber\\
&  \label{Delta theta g*tau>>1}%
\end{align}

By taking the square of Eq. (\ref{d theta /d tau}) one obtains%

\begin{equation}
\zeta^{2}+g\frac{\mathrm{d}\left(  \theta^{2}\right)  }{\mathrm{d}\tau}%
+g^{2}\theta^{2}=f^{2}\;, \label{ZitaSquareEq}%
\end{equation}
where the variable $\zeta$ is given by
\begin{equation}
\zeta(\tau)\equiv\frac{\mathrm{d}\theta}{\mathrm{d}\tau}\;.
\end{equation}

Expressing $\zeta(\tau)$ as a sum of a mean value and a deviation terms in a
similar manner to Eq. (\ref{theta(tau)}) yields%

\begin{equation}
\zeta\left(  \tau\right)  =\left\langle \zeta\left(  \tau\right)
\right\rangle +\Delta_{\zeta}\left(  \tau\right)  \;.
\end{equation}
To evaluate the variance of $\zeta\left(  \tau\right)  $, which is denoted as
$\left\langle \Delta_{\zeta}^{2}\left(  \tau\right)  \right\rangle $, in the
limit of relatively long times we employ Eqs. (\ref{ZitaSquareEq}),
(\ref{Delta theta g*tau>>1}), (\ref{f}), (\ref{bb}), (\ref{bbstar}) and get%

\begin{align}
\left\langle \Delta_{\zeta}^{2}\right\rangle  &  =\frac{g^{2}G\Theta_{\infty
0}^{2}}{\left\vert B_{\infty}\right\vert ^{2}}\frac{\omega_{0}}{\gamma}%
\frac{\left(  \omega_{p}-\omega_{0}\right)  ^{2}+\gamma\left(  \gamma
+g\omega_{0}\right)  }{\left(  \gamma+g\omega_{0}\right)  ^{2}+\left(
\omega_{p}-\omega_{0}\right)  ^{2}}\;.\nonumber\\
&
\end{align}

\section{ESCAPE RATE}

Escape from S to N states originates from a flux at point $\Theta=1$ (or
$\theta=1-\Theta_{\infty0}$) flowing from $\Theta<1$ to $\Theta>1$, or vise
versa for the case of escape from N to S states. Thus, the escape rate is
given by%

\begin{equation}
\Gamma=\omega_{0}\int_{0}^{\infty}\zeta f\left(  1-\Theta_{\infty0}%
,\zeta\right)  \mathrm{d}\zeta\;, \label{escape rate main}%
\end{equation}
where $f\left(  \theta,\zeta\right)  $ is the joint probability distribution
function of the random variables $\theta$ and $\zeta$. As was shown above, in
the limit where $g\tau\gg1$,\ the expectation values $\left\langle
\theta\right\rangle $ and $\left\langle \zeta\right\rangle $ vanish. In
general, $f\left(  \theta,\zeta\right)  $ is expected to represents a joint
normal distribution. Moreover, $\theta$ and $\zeta$ become statistically
independent as the expectation value $\left\langle \theta^{2}\right\rangle $
becomes time independent. This can be readily inferred from the following relation%

\begin{equation}
\left\langle \Delta_{\theta}\Delta_{\zeta}\right\rangle =\frac{1}{2}%
\frac{\mathrm{d}\left\langle \theta^{2}\right\rangle }{\mathrm{d}\tau
}-\left\langle \theta\right\rangle \left\langle \zeta\right\rangle \;.
\end{equation}

Thus, by applying the previous approximations one finds%

\begin{align}
\Gamma &  =\frac{\omega_{0}\exp\left[  -\frac{\left(  1-\Theta_{\infty
0}\right)  ^{2}}{2\left\langle \Delta_{\theta}^{2}\right\rangle }\right]
}{2\pi\sqrt{\left\langle \Delta_{\theta}^{2}\right\rangle \left\langle
\Delta_{\zeta}^{2}\right\rangle }}\int_{0}^{\infty}\zeta\exp\left(
-\frac{\zeta^{2}}{2\left\langle \Delta_{\zeta}^{2}\right\rangle }\right)
\mathrm{d}\zeta\;.\nonumber\\
&
\end{align}
Furthermore, by evaluating the integral, substituting instead of $G$ and
$\Theta_{\infty0}$ (given by Eqs. (\ref{G}) and (\ref{Theta_inf})
respectively), and using the notations%
\begin{equation}
C=\frac{1}{2}\frac{\left(  \gamma+g\omega_{0}\right)  ^{2}+\left(  \omega
_{p}-\omega_{0}\right)  ^{2}}{g\omega_{0}\left(  \gamma+g\omega_{0}\right)
}\;,
\end{equation}

\begin{equation}
\Gamma_{0}=\frac{\omega_{0}}{2\pi}\sqrt{\frac{g\left[  \left(  \omega
_{p}-\omega_{0}\right)  ^{2}+\gamma\left(  \gamma+g\omega_{0}\right)  \right]
}{\omega_{0}\left(  \gamma+g\omega_{0}\right)  }}\;, \label{Gamma0}%
\end{equation}
one gets%

\begin{equation}
\Gamma=\Gamma_{0}\exp\left[  -\frac{C\left(  U_{c}-U_{\infty}\right)  ^{2}%
}{U_{\infty}k_{B}T_{\mathrm{eff}}}\right]  \;, \label{escape total}%
\end{equation}
where%

\begin{equation}
U_{\infty}=\hbar\omega_{0}\left\vert B_{\infty}\right\vert ^{2}\;,
\end{equation}
is the energy stored in the resonator corresponding to the steady state
amplitude $B_{\infty}$, and%

\begin{equation}
U_{c}=\hbar\omega_{0}\left\vert B_{c}\right\vert ^{2}\;,
\end{equation}
is the mode energy corresponding to the critical amplitude $B_{c}$ at which
$\Theta_{\infty0}=1$, namely%

\begin{equation}
1=\Theta_{\infty0}=\frac{2\hslash\alpha\gamma_{2}\left\vert B_{c}\right\vert
^{2}}{gC\left(  T_{c}-T_{0}\right)  }\;.
\end{equation}
Note that typically in our NbN devices \cite{dynamics} $\gamma/g\omega
_{0}\simeq10^{-2}$. Thus, by assuming the limit $\gamma/g\omega_{0}\ll1$, and
the resonance case $\omega_{p}=\omega_{0}$, the above expression appearing in
Eq. (\ref{escape total}) reduces into%

\begin{equation}
\Gamma=\frac{\sqrt{g\omega_{0}\gamma}}{2\pi}\exp\left[  -\frac{1}{2}%
\frac{\left(  U_{c}-U_{\infty}\right)  ^{2}}{U_{\infty}k_{B}T_{\mathrm{eff}}%
}\right]  \;. \label{eta for omega=omega_p}%
\end{equation}

\section{STOCHASTIC RESONANCE}

In order to examine experimentally the escape rate expression derived in Eq.
(\ref{escape total}), we employed stochastic resonance technique. Basically,
stochastic resonance phenomenon demonstrates how a weak periodic signal,
applied to a nonlinear metastable system, can be amplified at the system
output with the aid of certain amount of zero-mean Gaussian white noise. The
amplification of the signal occurs when a resonant cooperation is established
between the small periodic signal and the white noise entering the system. In
general, such a coherent interaction between the signal and the noise occurs
when the angular frequency $\Omega$ of the signal, which periodically
modulates the double-well potential of the system, becomes comparable to the
escape rate of the metastable states in the presence of the white noise.

\begin{figure}
[ptb]
\begin{center}
\includegraphics[
height=2.4941in,
width=3.4134in
]%
{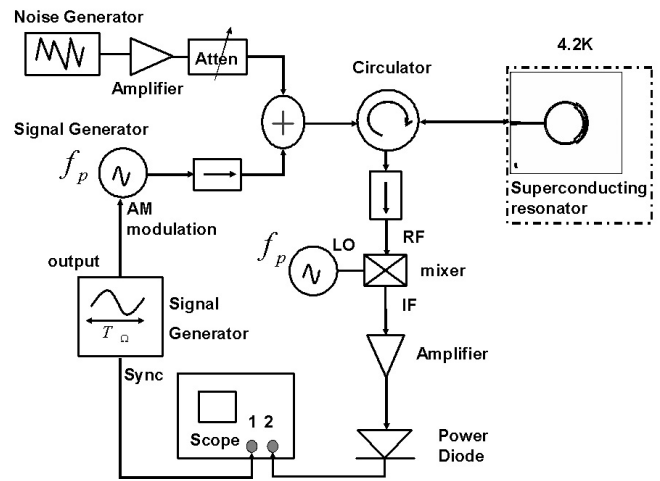}%
\caption{Schematic block diagram of the experimental setup used in order to
measure stochastic resonance. The microwave signal generator and the local
oscillator at frequency $f_{p}$ were phase locked. The layout of the resonator
is shown at the top-right corner.}%
\label{srsetup}%
\end{center}
\end{figure}

The stripline center-layer layout of the NbN resonator employed in the
measurements is shown at the top-right corner of Fig. \ref{srsetup}. The
resonator was dc-magnetron sputtered on a $34%
\operatorname{mm}%
$X $30%
\operatorname{mm}%
$X $1%
\operatorname{mm}%
$ Sapphire substrate in an ambient gas mixture of Ar/N$_{2}$ at room
temperature. The resonator was patterned using standard optical lithography
and ion-milling. The resonator thickness was set to $2200%
\operatorname{\text{\AA}}%
$. Additional fabrication process parameters are listed in Ref.
\cite{observation}. Whereas modeling and characterization of these nonlinear
resonators are elaborated in Ref. \cite{dynamics}.

\begin{figure}
[ptb]
\begin{center}
\includegraphics[
height=2.7129in,
width=3.2707in
]%
{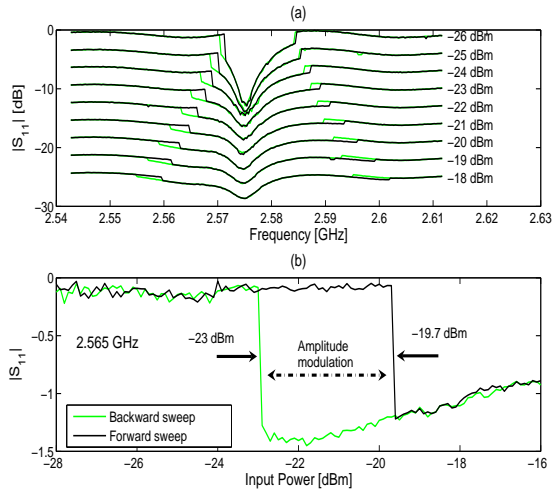}%
\caption{(Color online). (a) Forward and backward frequency sweeps applied to
the first mode of the resonator at $\sim2.575\operatorname{GHz}.$ The sweeps
exhibit hysteresis loops at both sides of the resonance line shape. The plots
corresponding to different input powers were shifted by a vertical offset for
clarity. (b) Reflected power hysteresis measured at a constant angular
frequency of $\omega_{p}=2\pi\cdot2.565\operatorname{GHz}$ which resides
within the left-side metastable region of the resonance. For both plots the
black (dark) line represents a forward sweep whereas the green (light) line
represents a backward sweep.}%
\label{hysteresis}%
\end{center}
\end{figure}

The metastable states of the system in our case, are manifested by the
occurrence of jumps in the resonance line shape of the resonator. In Fig.
\ref{hysteresis} (a) we show a reflection parameter measurement of the first
resonance mode of the resonator $f_{0}=\omega_{0}/2\pi\simeq2.575%
\operatorname{GHz}%
$, which exhibits two frequency hysteresis loops forming at both sides of the
resonant curve as the microwave frequency is swept in the forward and backward
directions. In Fig. \ref{hysteresis} (b) on the other hand, we show a
reflected power hysteretic behavior measured at a constant frequency
$f_{p}=\omega_{p}/2\pi=2.565%
\operatorname{GHz}%
$ which falls within the instable region of the fundamental mode, as the input
power is swept up and down. Thus, in order to drive our resonators into
metastability, we have applied a coherent microwave signal at frequency
$f_{p}$ and input power $P_{0}=-21.5$ dBm. Moreover, in order to tune the
resonator into stochastic resonance condition, we have applied a small
sinusoidal forcing to the system in the form of amplitude modulation, and
injected a thermal white noise with an adjustable intensity to the resonator port.%

A schematic illustration of the stochastic resonance measurement setup used is
depicted in Fig. \ref{srsetup}. A continuous microwave signal at frequency
$f_{p}$ is amplitude modulated at frequency $f_{\Omega}=\Omega/2\pi=1%
\operatorname{kHz}%
$. The modulated signal which effectively modulates the height of the
potential barrier is combined with a white noise and fed to the
superconducting resonator. The reflected signal off the resonator on the other
hand, is mixed with a local oscillator of frequency $f_{p}$ and measured in
the time domain using an oscilloscope. Additional information regarding
stochastic resonance phenomenon measured in these nonlinear superconducting
resonators is summarized in Ref. \cite{SRB}.%

\subsection{Escape Rate Measurement}

\begin{figure}
[ptb]
\begin{center}
\includegraphics[
height=2.655in,
width=3.4411in
]%
{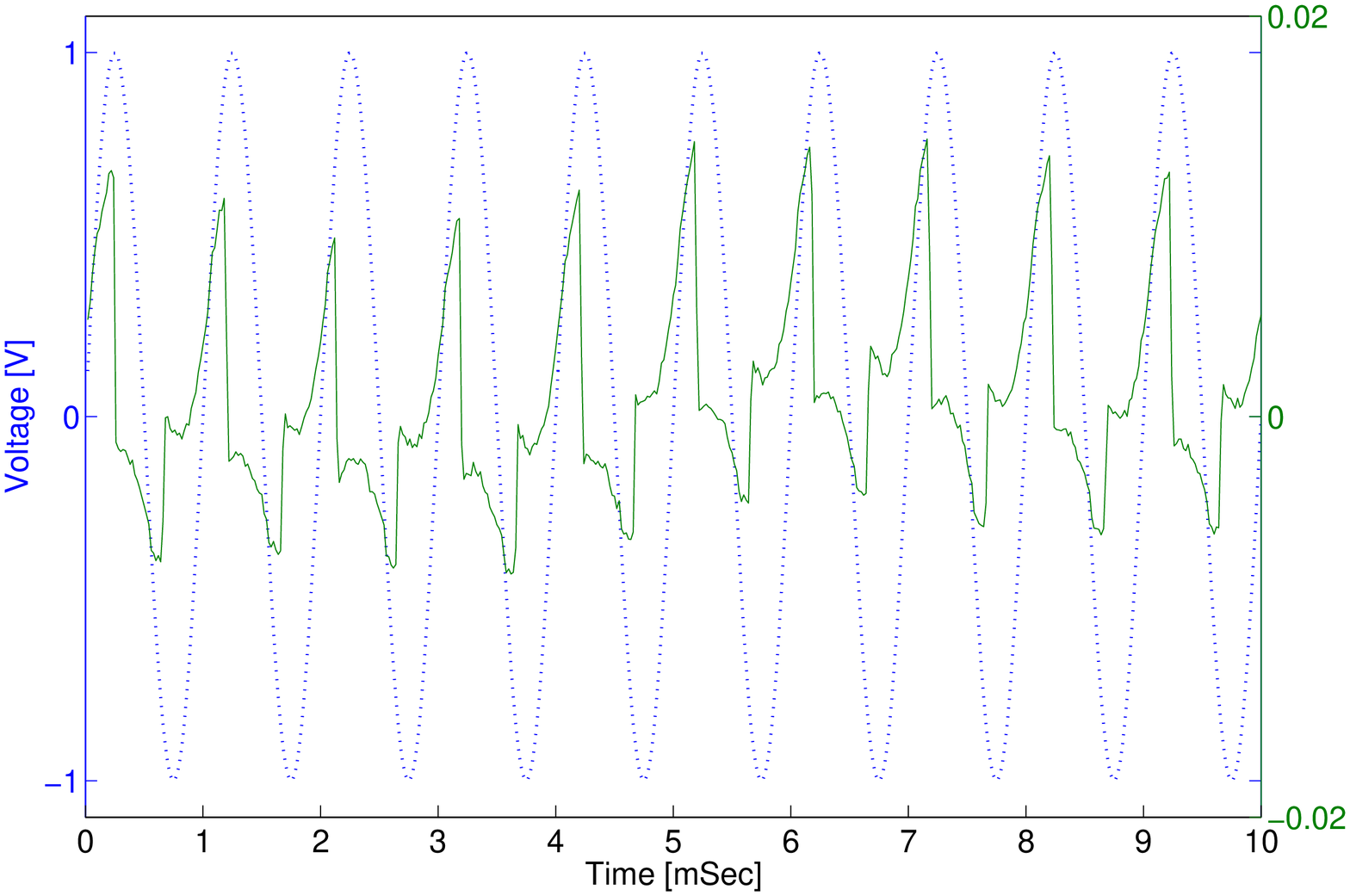}%
\caption{(Color Online). A typical snapshot of the time domain as the
resonator is tuned into stochastic resonance condition. The solid (green) line
represents the reflected modulated signal, corresponding to ten modulation
cycles out of 5000 employed in the analysis. The dotted (blue) sinusoidal line
represents the modulation signal applied to the microwave signal generator. }%
\label{TimeDomain}%
\end{center}
\end{figure}

At stochastic resonance condition, the lifetime of the metastable states
becomes approximately equal to half the modulation period. Thus, assuming that
the system has two metastable states designated by $S_{u}$ and $S_{d}$, one
obtains at this condition, one metastable state escape event each half time
cycle. This is shown in Fig. \ref{TimeDomain}, which shows a typical result
taken in the time domain at stochastic resonance conditions, where the jumps
appearing in the output signal correspond to alternating $S_{u}\rightarrow
S_{d}$ and $S_{d}\rightarrow S_{u}$ transitions.

The blue dotted line shows the amplitude modulation signal, which modulates
the escape rates $\Gamma_{1}$ and $\Gamma_{2}$ of the transitions
$S_{d}\rightarrow S_{u}$ and $S_{u}\rightarrow S_{d}$ respectively. Near the
minimum (maximum) points of the amplitude modulation signal the rate
$\Gamma_{1}$ ($\Gamma_{2}$) obtains its largest value, which is denoted as
$\Gamma_{\mathrm{m1}}$ ($\Gamma_{\mathrm{m2}}$). Let $\tau_{1}$ ($\tau_{2}$)
be the difference between the time of the transition $S_{d}\rightarrow S_{u}$
($S_{u}\rightarrow S_{d}$) and the time at which the corresponding escape rate
obtains its largest value, namely the time at which $\Gamma_{1}=\Gamma
_{\mathrm{m1}}$ ($\Gamma_{2}=\Gamma_{\mathrm{m2}}$). The probability density
of the random variable $\tau_{1}$ ($\tau_{2}$) is denoted by $f_{1}\left(
\tau_{1}\right)  $ ($f_{2}\left(  \tau_{2}\right)  $).

An estimate for the escape rates $\Gamma_{\mathrm{m1}}$ and $\Gamma
_{\mathrm{m2}}$ could be obtained by measuring the probability densities
$f_{1}\left(  \tau_{1}\right)  $ and $f_{2}\left(  \tau_{2}\right)  $. As can
be seen from Eq. (\ref{gamma0fit}) in the appendix, $\Gamma_{\mathrm{m1}}$ and
$\Gamma_{\mathrm{m2}}$ can be estimated from the expectation value and the
variance of the corresponding random variables $\tau_{1}$ and $\tau_{2}$.
However, a more accurate value of the prefactor $\Gamma_{\mathrm{m}}$ can be
obtained by invoking Eq. (\ref{lambda(tau)}) and using the measured
probability density function $f\left(  \tau\right)  $.

\begin{figure}
[ptb]
\begin{center}
\includegraphics[
height=2.7709in,
width=3.4411in
]%
{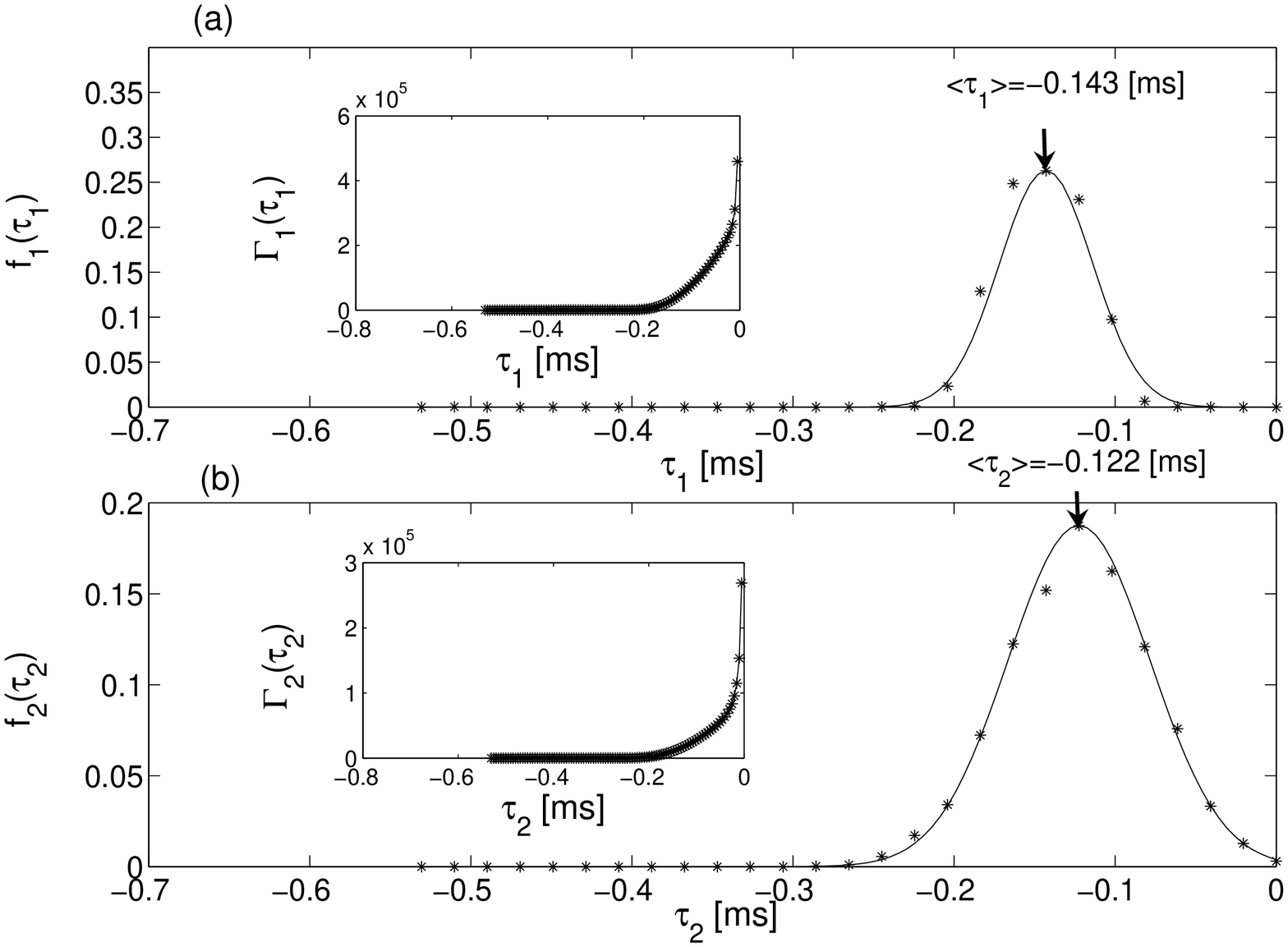}%
\caption{Gaussian probability density functions $f_{1}(\tau_{1})$ and
$f_{2}(\tau_{2})$ fitted to the experimental data which correspond to the
$S_{d}\rightarrow S_{u}$ transition in panel (a) and to the $S_{u}\rightarrow
S_{d}$ transition in panel (b). The escape rates $\Gamma_{1}$ and $\Gamma_{2}$
associated with both transitions is plotted in the insets of panels (a) and
(b) respectively as a function of the random time variables $\tau_{1}$ and
$\tau_{2}$ according to Eq. (\ref{lambda(tau)}).}%
\label{f_tau2transitions}%
\end{center}
\end{figure}

In Fig. \ref{f_tau2transitions} (a) and (b) we show the measured probability
densities $f_{1}\left(  \tau_{1}\right)  $ and $f_{2}\left(  \tau_{2}\right)
$ derived from $5000$ modulation cycles sampled in the time domain. The solid
line in both panels represent a Gaussian function fitted to the measured
probability density in each case. The transition rate $\Gamma_{1}$
($\Gamma_{2}$) as a function of the random variable $\tau_{1}$ ($\tau_{2}$),
which is found using Eq. (\ref{lambda(tau)}) and the Gaussian fit, is shown in
the inset of Fig. \ref{f_tau2transitions} (a) [Fig. \ref{f_tau2transitions}
(b)]. These plots yield also the values $\Gamma_{\mathrm{m1}}\simeq$
$4.6\cdot10^{5\text{ }}%
\operatorname{Hz}%
$ and $\Gamma_{\mathrm{m2}}\simeq$ $2.7\cdot10^{5\text{ }}%
\operatorname{Hz}%
$ for the transitions $S_{d}\rightarrow S_{u}$ and $S_{u}\rightarrow S_{d}$
respectively.%
%

\subsection{Discussion}

In order to obtain theoretical estimates for the escape rates $\Gamma
_{\mathrm{m1}}$ and $\Gamma_{\mathrm{m2}}$ corresponding to the $S_{d}%
\rightarrow S_{u}$ and $S_{u}\rightarrow S_{d}$ transitions respectively we
rewrite Eq. (\ref{escape total}) in terms of the feedline input power $P_{in}$
at the extremum of the AM modulation, and the power difference $\Delta
P_{in}\equiv P_{c}-P_{in}$, where $P_{c}$ is the critical input power, which
is proportional to $U_{c}$,%

\begin{equation}
\Gamma_{\mathrm{m}}=\Gamma_{0}\exp\left[  -C\left(  \frac{\Delta P_{in}%
}{P_{in}}\right)  ^{2}\frac{U_{\infty}}{k_{B}T_{\mathrm{eff}}}\right]  \;,
\label{GammaPin}%
\end{equation}
where, according to Eq. (\ref{dB/dt}), the following holds%
\begin{equation}
U_{\infty}=\frac{P_{in}\left(  1-\left\vert r\right\vert ^{2}\right)
}{2\gamma}\;. \label{UinfPin}%
\end{equation}
\qquad%

\begin{table}[ptb]
\caption{Calculated and Measured Model Parameters}
\label{Model_Params}
\renewcommand{\arraystretch}{1.5} \begin{ruledtabular}
\begin{tabular}{cccccc}
& $ S_{d}  \rightarrow S_{u} $ & $ S_{u} \rightarrow S_{d} $ \\
\hline\hline$ g ~[10^{-3}]$ & $1.56$ & $1.56$ \\\hline$ \gamma[
\mathrm{MHz}
]$ & $37.6 $ & $18.6$ \\\hline$P_{c}~[$dBm$]$ & $-23$ & $-19.6$ \\\hline
$\Delta P_{in} [10^{-6}
\mathrm{W}
]$ & $0.12$ & $1.2$ \\\hline$\left\vert r \right\vert^2 $ &
$ 0.55 $ & $ 0.8 $ \\\hline$ k_{B} T_{\textrm{eff}} ~[\textrm{f}
\mathrm{W}/\mathrm{Hz}]$ & $1.4$ & $1.4$ \\\hline$\Gamma_{0} [
\mathrm{Hz}
]$ & $ 8 \cdot10^6 $ & $ 8.3 \cdot10^6 $ \\\hline$\Gamma_{\textrm{m}}
[
\mathrm{Hz}
]$ ~(calc.) & $ 7.8  \cdot10^6 $ & $ 1.9 \cdot10^6 $ \\\hline$\Gamma
_{\textrm{m}} [
\mathrm{Hz}
]$ ~(meas.) & $ 4.6  \cdot10^5 $ & $ 2.7 \cdot10^5 $ \\
\end{tabular}
\end{ruledtabular}\end{table}%

Estimates for the model parameters corresponding to both $S_{d}\rightarrow
S_{u}$ and $S_{u}\rightarrow S_{d}$ transitions are summarized in Table 1.
Estimates for the coupling parameter $\gamma$ corresponding to the $S_{u}$ and
the $S_{d}$ states have been extracted indirectly by fitting Eq. (\ref{r}) to
the measured $r$ vs. $\omega_{p}$ curves in the vicinity of the resonance
\cite{cc}. Whereas, the cooling parameter $g$, which is defined in Eq.
(\ref{g}), has been estimated using experimentally measured material
properties of NbN \cite{Bol,Nonequilibrium,Use}, yielding the value
$g\simeq1.56\cdot10^{-3}$ (see Refs. \cite{dynamics,Segev long}). Employing
these estimates together with the experimental values of $P_{c}$, $\Delta
P_{in}$, $r$, and substituting in Eq. (\ref{GammaPin}) yield a rough estimate
for the escape rates $\Gamma_{\mathrm{m1}}$ $\simeq7.8\cdot10^{6}%
\operatorname{Hz}%
$ and $\Gamma_{\mathrm{m2}}\simeq1.9\cdot10^{6}%
\operatorname{Hz}%
$ for the $S_{d}\rightarrow S_{u}$ and the $S_{u}\rightarrow S_{d}$
transitions respectively.

The discrepancy in the values of the escape rates obtained using the
theoretical model as opposed to the ones extracted from the experimental data
(by about an order of magnitude) can be attributed most likely to the
accumulated errors in the estimated values of the model parameters which have
been evaluated indirectly. For example, the $g$ parameter depends among others
on the geometry of the hot spot and the thermal properties of the deposited
NbN film, which are not known precisely. Moreover, inaccuracies in the
coupling factor $\gamma$ may result due to some approximations, which were
employed in the fitting procedure \cite{cc}.

\section{SUMMARY}

In conclusion, a noise-activated escape rate expression was derived for the
case of a nonlinear superconducting resonator having a local-thermal
instability. Moreover, stochastic resonance measurements were exploited to
experimentally determine the escape rate. A partial agreement is found between
the theoretical and the experimental results.

\appendix

\section{Transition Lifetime}

Consider a system which has in general two metastable states designated by
$S_{a}$ and $S_{b}$ and assume that at time $t=-t_{0}$ the system is in state
$S_{b}$, where $t_{0}>0$. The transition rate $\Gamma$ of the process
$S_{b}\rightarrow S_{a}$ depends on an externally applied time varying
parameter $p\left(  t\right)  $. Further assume that for $p$ close to some
fixed value $p_{\mathrm{m}}$ the transition rate is given approximately by%

\begin{equation}
\Gamma\left(  p\right)  =\Gamma_{\mathrm{m}}\exp\left(  -\kappa^{2}%
\frac{p-p_{\mathrm{m}}}{p_{\mathrm{m}}}\right)  \;, \label{gammaP}%
\end{equation}
where both $\Gamma_{\mathrm{m}}$ and $\kappa$ are positive constants.

The probability distribution function $F\left(  \tau\right)  $ for a
transition of the kind $S_{b}\rightarrow S_{a}$ to take place within the time
interval $\left(  -t_{0},\tau\right)  $ is given by
\begin{equation}
F\left(  \tau\right)  =\int_{-t_{0}}^{\tau}f\left(  t\right)  \mathrm{d}t\;,
\label{Ftau}%
\end{equation}
where $f\left(  \tau\right)  $ is the corresponding probability density. By
definition, the following holds%

\begin{equation}
\frac{f\left(  \tau\right)  }{1-F\left(  \tau\right)  }=\Gamma\left[  p\left(
\tau\right)  \right]  \;. \label{lambda(tau)}%
\end{equation}
The initial condition $F\left(  -t_{0}\right)  =0$ and Eq. (\ref{lambda(tau)}) yield%

\begin{equation}
f\left(  \tau\right)  =\Gamma\left[  p\left(  \tau\right)  \right]
\exp\left(  -\int_{-t_{0}}^{\tau}\Gamma\left[  p\left(  t\right)  \right]
\mathrm{d}t\right)  \;.
\end{equation}

Further assume the case where at time $t=0$ the function $p\left(  t\right)  $
obtains a local minimum $p\left(  0\right)  =p_{\mathrm{m}}$. Near $t=0$ one has%

\begin{equation}
p\left(  t\right)  =p_{\mathrm{m}}\left(  1+\Omega^{2}t^{2}\right)  +O\left(
t^{3}\right)  \;.
\end{equation}
Thus, in the vicinity of $t=0$ Eq. (\ref{gammaP}) becomes%

\begin{equation}
\Gamma\left(  t\right)  =\Gamma_{\mathrm{m}}\exp\left(  -\kappa^{2}\Omega
^{2}t^{2}\right)  \;, \label{GammaTime}%
\end{equation}
and the following holds%

\begin{align}
f\left(  \tau\right)   &  =\Gamma_{\mathrm{m}}\exp\left(  -\kappa^{2}%
\Omega^{2}\tau^{2}-\sqrt{\pi}\frac{\Gamma_{\mathrm{m}}}{\kappa\Omega}%
\frac{\mathrm{erf}\left(  \kappa\Omega\tau\right)  +\mathrm{erf}\left(
\kappa\Omega t_{0}\right)  }{2}\right)  \;.\nonumber\\
&
\end{align}
Keeping terms up to second order in $\kappa\Omega\tau$ and assuming the case
where
\begin{equation}
\left(  -\kappa\Omega t_{0}+\frac{\Gamma_{\mathrm{m}}}{2\kappa\Omega}\right)
^{2}\gg1\;,
\end{equation}
allow approximating the probability density $f\left(  \tau\right)  $ by%

\begin{equation}
f\left(  \tau\right)  =\frac{\Omega\kappa}{\sqrt{\pi}}\exp\left[  -\kappa
^{2}\Omega^{2}\left(  \tau+\frac{\Gamma_{\mathrm{m}}}{2\kappa^{2}\Omega^{2}%
}\right)  ^{2}\right]  \;. \label{f(tau)}%
\end{equation}
In this approximation the random variable $\tau$ has a normal distribution
function with a mean value%

\begin{equation}
\mu_{\tau}=-\frac{\Gamma_{\mathrm{m}}}{2\kappa^{2}\Omega^{2}}\;,
\end{equation}
and a variance%

\begin{equation}
\sigma_{\tau}^{2}=\frac{1}{2\kappa^{2}\Omega^{2}}\;.
\end{equation}
Whereas, the parameters $\Gamma_{\mathrm{m}}$ and $\kappa,$ are given by%

\begin{equation}
\Gamma_{\mathrm{m}}=-\frac{\mu_{\tau}}{\sigma_{\tau}^{2}}\;, \label{gamma0fit}%
\end{equation}
and%

\begin{equation}
\kappa^{2}=\frac{1}{2\sigma_{\tau}^{2}\Omega^{2}}\;. \label{alphaSqfit}%
\end{equation}


\section*{Acknowledgment}

This work was supported by the German Israel Foundation under grant
1-2038.1114.07, the Israel Science Foundation under grant 1380021, the Deborah
Foundation, the Poznanski Foundation, and MAFAT.

\bibliographystyle{plain}
\bibliography{apssamp}

\end{document}